\documentclass[aps,prl,twocolumn,epsf,epsfig,amsmath]{revtex4}[floatfix]
\usepackage{latexsym}
\usepackage{hyperref}
\usepackage{times}
\usepackage{graphicx}
\usepackage{amsmath}
\usepackage{multirow} 
\usepackage{dcolumn}
\usepackage{bm}
\usepackage{ textcomp }
\usepackage{color}
\usepackage{amssymb}
\usepackage{xcolor}
\usepackage{easyReview}
\usepackage{mathdots}
\usepackage{float}
\usepackage{lineno}
\usepackage{todonotes}
\usepackage{array}
\usepackage{multirow}
\usepackage{makecell}
\usepackage[title]{appendix}
\newcommand{\beq}{\begin{eqnarray}}
\newcommand{\eeq}{\end{eqnarray}}
\usepackage[utf8]{inputenc}
\usepackage{hyperref}
\usepackage{amsmath, amsfonts, amssymb}
\usepackage{bbm}
\usepackage{physics}
\usepackage{graphicx}
\usepackage{amsthm}

\begin{document}
\title{Dynamical Spectral Weight Transfer in the Orbital HK Model}
\author{Gaurav Tenkila, Jinchao Zhao, Philip W. Phillips}

\affiliation{Department of Physics and The Anthony J Leggett Institute for Condensed Matter Theory, University of Illinois at Urbana-Champaign, Urbana, IL 61801, USA}

\begin{abstract}
We compute explicitly the low-energy spectral weight (LESW) in the lower Hubbard band using the orbital Hatsugai-Kohmoto (OHK) model.  We show that the dynamical mixing enters the LESW with a universal slope independent of the number of orbitals.  This independence on orbital number corroborates the rapid convergence of the OHK model to the Hubbard model. As a result, OHK is an effective simulator of Hubbard physics.
\end{abstract}
\date{\today}

\maketitle

The Landau one-to-one correspondence guarantees that anytime a single electron is removed from a Fermi liquid, just a single empty k-state appears above the chemical potential. As a result, counting electrons completely exhausts the spectral weight in a Fermi liquid. However, for doped Mott insulators, this sum rule fails as shown originally by Harris and Lange\cite{hl} in the Hubbard model and confirmed experimentally\cite{chen}.  The root cause is the bifurcation of the spectral weight per k-state into lower and upper Hubbard bands which leads experimentally to a breakdown 1-1 correspondence in Fermi liquid theory.  As a result of this bifurcation, every hole doped into a Mott insulator creates two empty states immediately above the chemical potential. 
 There are course dynamical corrections arising from the kinetic mixing between the bands.  Consequently, the low-energy spectral weight in doped Mott insulators increases as
\beq
\Lambda(x)=2x+f(t,x,U)>2x,
\eeq
where $f(t,x,U)$ is determined by the projected kinetic energy\cite{hl,sawatzky,eskes} in the lower band.
The inequality stems from the fact that the dynamical correction is inherently positive as it is mediated by double occupancy which necessarily leads to empty states in the lower band.  The occupied (below the chemical potential) part of the lower band has a spectral weight of $1-x$.   As a consequence, the total weight of the lower band is $1+x+f(t,U)>1+x$.  However, only $1+x$ electrons can fit into the lower band.  Consequently, counting electrons does not exhaust the low-energy spectral weight in a doped Mott insulator, and degrees of freedom that have no interpretation in terms of standard fermionic quasiparticles populate the band.  This is not surprising as Mottness is inherently about the breakdown of the single-particle concept.  

Typically, perturbative calculations\cite{hl,sawatzky,lcp}, phenomenology\cite{schmalian} or numerical methods\cite{eskes} are employed to determine the mixing function, $f(t,x,U)$.    We show here that it can be determined precisely from a solvable model for a doped Mott insulator.  This constitutes the first time the low-energy spectral weight in a doped Mott insulator has been calculated exactly in any soluble model. The mixing contribution between the bands implies that the spectral weight depends on the parameters in the Hamiltonian, quite different from a Fermi liquid in which it is strictly the size of the Hilbert space that determines low-energy degrees of freedom. 

\section{Model and Setup}

The starting point for our analysis is the orbital version\cite{barry,OHK} of
the Hatsugai-Kohmoto\cite{hk} (HK) model, given by the following tight-binding Hamiltonian on a square lattice
\begin{equation}
    \begin{aligned}
 H_{\text{OHK}}=& \sum_{\bf k,\alpha,\alpha',\sigma}g_{\alpha,\alpha'}({\bf k})c_{{\bf k}\alpha\sigma}^\dagger c_{{\bf k}\alpha'\sigma}-\mu\sum_{{\bf k},\alpha}n_{{\bf k}\alpha,\sigma} \\&+\sum_{{\bf k},\alpha,\alpha'} U_{\alpha,\alpha'} n_{{\bf k}\alpha\uparrow} n_{{\bf k}\alpha'\downarrow},
 \label{ohk}
 \end{aligned}
\end{equation}
with the band index given by $k$ and $\alpha(\beta)$ label the orbital indices.
While the band-HK model lacks DSWT, not so in the orbital extension which we have shown numerically\cite{OHK} contains such a contribution in excellent agreement with simulations\cite{sawatzky} on the 1D Hubbard model.  As we also pointed out previously\cite{OHK}, the orbital Hatsugai-Kohmoto (OHK) model converges rapidly to the Hubbard model regardless of the spatial dimension with quantitative agreement for double occupancy achieved already at two or four-orbital extensions.  For the first section of this note, we will set the number of orbitals $N_\alpha=2$ and employ the dispersion matrix,
\beq
g = -2t\eta({\bf k}) \sigma_x,
\eeq
where $\sigma_1$ is the first Pauli matrix acting on the orbital indices, $t$ is the hopping amplitude and $\eta(\bf k)$ is the single-particle dispersion given by
\begin{equation}
\eta({\bf k}) =2 \ \text{cos}\left(\frac{k_x}2\right)\text{cos}\left(\frac{k_y}2\right). 
\end{equation}
We will later explore how the results vary with orbital number and with the lattice dimensionality. It is to be noted that there is a minor difference in convention for stating the dispersion here. Mai et al. \cite{OHK} use a reduced Brillouin zone convention whereas we use the standard Brillouin zone with the explicit scaling for $\mathbf{k}$.  The model thus stated is diagonal in $\bf k$ space and thus can be exactly diagonalized by employing a complete basis in Fock space. We can thus exactly compute observables on each $\bf k$ site and proceed to the continuum limit with the substitution
\begin{equation}
    \sum_{{\bf k}\in \text{BZ}} \longrightarrow \frac{1}{4\pi^2} \int_{{\bf k}\in \text{BZ}}.
\end{equation}
With malice aforethought, we assume that we are in the strongly interacting limit with the bound $U>16\sqrt{3}t$.

\begin{figure}
    \centering
    \includegraphics[scale=0.35]{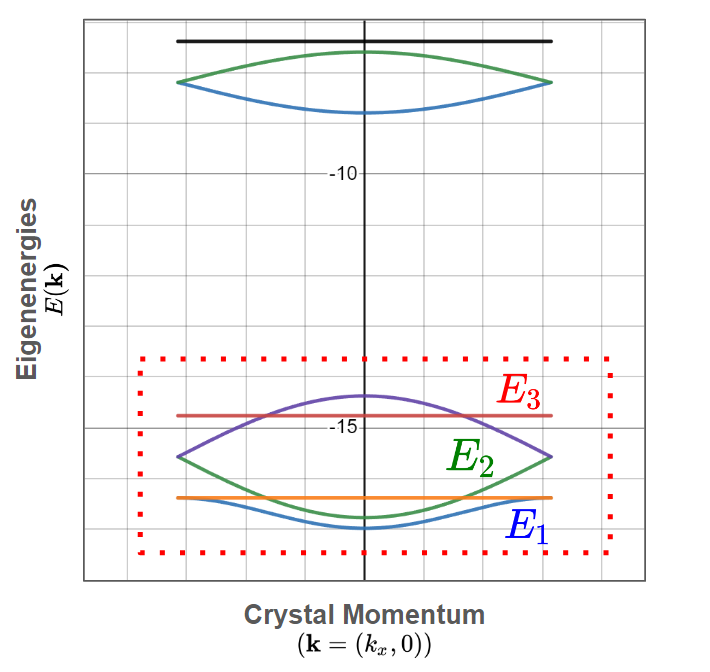}
    \caption{The 8 eigenbands for 2-orbital HK along a slice of the Brillouin zone. The box indicates the low energy sector of the model.}
    \label{fig:eigenbands}
\end{figure}

The Hilbert space of the system can be decomposed into a direct product over $\bf k$ states as $\mathcal H \longrightarrow \bigotimes_k \mathcal{H}_k$, where each $\mathcal{H}_k$ has 16 basis states. This can be exactly diagonalized to yield 8 distinct eigenbands as shown in Fig.  \ref{fig:eigenbands}. Since we are interested in the low-energy spectral weight and for hole doping, we only need to consider
the lowest 5 eigenbands, the boxed region in Fig. \ref{fig:eigenbands}.

The model also conserves particle number, allowing us to decompose the Fock space at each $\bf k$ state into fixed particle number sectors. Since we will be considering electron doping from half filling for the 2-orbital HK model, the relevant sectors are $N=2,3$ and $4$. The states can be labelled by a binary index, $\ket{n_{\bf{k}1\downarrow} \ n_{\bf{k}1\uparrow} \ n_{\bf{k}2\downarrow} \ n_{\bf{k}2\uparrow}}$. Table. \ref{tab:hkdecomp} enumerates the basis states for the different particle sectors, along with the associated eigenvalues. 

\begin{table*}
    \centering
    \begin{tabular}{|c|c|c|}
        \hline
        Particle Sector & Eigenvalue & Eigenvectors \\ \hline
        $N=2$ & $E_1=\frac12 \left(U-4sU-\sqrt{U^2+(8t\eta)^2}\right)$ & \makecell {$\frac12 \sqrt{1-\frac{U^2}{\sqrt{U^2+16(2t\eta)^2}}}(\ket{0011}+\ket{1100})$ \\ $+\frac12 \sqrt{1+\frac{U^2}{\sqrt{U^2+16(2t\eta)^2}}}(\ket{1001} + \ket{0110})$} \\ \hline
        $N=3$ & $E_2=U-3sU-2t\eta$ & \makecell{$\frac{1}{\sqrt2}(\ket{1101}+\ket{0111})$, \\ $\frac{1}{\sqrt2}(\ket{1110}+\ket{1011})$}    \\ \hline
        $N=4$ & $E_3 = 2U-4sU$ & $\ket{1111}$ \\ \hline
    \end{tabular}
    \caption{Block decomposition and the energy levels of the 2-orbital HK Hamiltonian}
    \label{tab:hkdecomp}
\end{table*}

\section{Occupancy}
\begin{figure}
    \centering
    \includegraphics[scale=0.27]{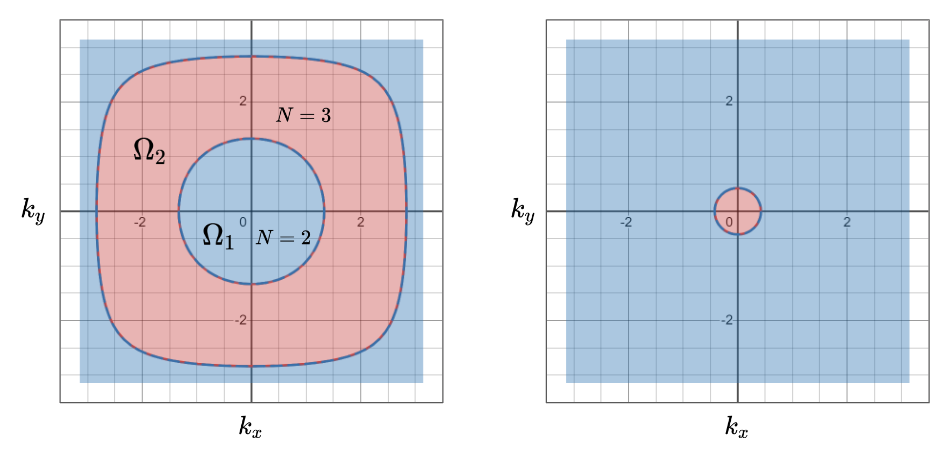}
    \caption{Partitioning of the Brillouin Zone into eigen-regions indexed by the occupation eigenvalue. The figure on the left represents the partitioning under electron doping for a system with $t/U>E^\ast$, whereas the figure on the right represents $t/U<E^\ast$.}
    \label{fig:BZpartition}
\end{figure}

We first need to determine the total occupancy of the system as a function of the Hamiltonian parameters (in particular, the chemical potential $\mu$). The deviation from half filling gives us the dopant concentration. The total ground state occupancy is given by
\begin{equation}
n(\mu) = \frac{1}{4\pi^2}\int d{\bf k} \frac12\sum_{\alpha, \sigma} \bra{\psi_0({\bf k})}c_{{\bf k}\alpha\sigma }^\dagger c_{{\bf k}\alpha\sigma } \ket{\psi_0(\bf k)},
\end{equation}
where $\ket{\psi_0(\bf k)}$ is the ground state of the system with momentum  $\bf k$. We proceed by identifying the lowest eigenenergies as a function of $\bf k$ and noting that the bands partition the Brillouin zone. Fig. \ref{fig:BZpartition} depicts the partitioning between the $N=3$ and $N=2$ bands under electron doping for two different values of $t/U$. It is important to note that the partitioning is non-trivial for general values of $t/U$. We obtain a circular partition that is analytically tractable only under the strongly interacting criterion $t/U < \frac{1}{16\sqrt3} = r^\ast$.

We define a dimensionless parameter $s$ as $\mu=sU$. Note that for $0.5\leq s\leq1$, the ground state energies for all $\bf k$ are either $E_1$ or $E_2$. For $s\geq1$, the ground states energies are $E_2$ or $E_3$. We will restrict ourselves to the former case, since the procedure for the latter is analogous.
 The two bands partition the Brillouin zone into $\Omega_1$ and $\Omega_2$, where the ground state is either given by $E_1$ or $E_2$, respectively. This will lead to a two-integral contribution,
\begin{equation}
    \begin{aligned}
    8\pi^2n(s) = &\int_{{\bf k}\in\Omega_1} d{\bf k} \bra{E_1(\bf k)}c_{\bf k}^\dagger c_{\bf k} \ket{E_1(\bf k)}\\
    &+\int_{{\bf k}\in\Omega_2} d{\bf k} \bra{E_2(\bf k)}c_{\bf k}^\dagger c_{\bf k} \ket{E_2(\bf k)},
    \end{aligned}
\end{equation}
to the occupancy.
The expectation values in fact do not depend on $\bf k$ and are simply equal to $2$, $3$ and $4$ for $E_1, E_2$ and $E_3$ respectively. Thus, the problem is reduced to finding the areas of the regions partitioned by the implicitly defined closed curve $E_1({\bf k}) = E_2({\bf k})$:
\begin{align}\label{eqn:occ_general}
    n(s) &= \frac{1}{4\pi^2}\left(|\Omega_1|+\frac{3}{2}|\Omega_2| \right) \nonumber \\
         &= \frac{4\pi^2-|\Omega_2|+1.5|\Omega_2|}{4\pi^2} \nonumber \\
         &= 1 + \frac{1}{8\pi^2}|\Omega_2|.
\end{align}

The area of this region, in general, does not have an analytical solution. However, if we restrict to the weak doping limit, where the two bands just begin to intersect, we can expand the cosine around the origin. This gives us a circular region where the ground state is $E_2(\bf k)$, with a radius given by,
\beq
    R^2(s) =  \frac{\left(4t-\frac{64t^{2}}{U}+U\left(s-1\right)\right)}{\frac{t}2-\frac{16t^{2}}{U}}.
\eeq
Note that the validity of the above formula is only for a small range around $d$ where $E_1 (\bf k)$ and $E_2 (\bf k)$ intersect:
\begin{gather}
    s\in (s^*, s^*+\epsilon) \\
    s^* = \frac{1}{2}-\frac{4t}{U}+\frac12\sqrt{1+\frac{256t^2}{U^2}}.
\end{gather}
Thus, in the low doping regime, 
\begin{equation}
    n(s) = 1+\frac{R^2(s)}{8\pi}
\end{equation}
defines the occupancy.

\section{Spectral Weight}
We now relate the occupancy to the spectral function. 
As is well known, the imaginary part of the Green function,
\begin{gather}
    A(\omega) = \sum_{\alpha, \sigma}\int d{\bf k}\frac1\pi \text{Im } G_{\alpha, \sigma}({\bf k}, \omega) \\
    G_{\alpha, \sigma}({\bf k}, \omega) = \sum_m \frac{|\bra{\psi_m({\bf k})}c_{{\bf k \alpha \sigma }}\ket{\psi_0({\bf k})}|^2}{\omega - E_n + E_0 +i\eta^+},
\end{gather}
defines the spectral function.
Taking the continuum limit and summing over the orbitals yields our working expression,
\begin{multline}
    A(\omega) = \frac{1}{4\pi^2}\int d{\bf k} \sum_{m}|\bra{\psi_m({\bf k})}c_{{\bf k \alpha \sigma }}\ket{\psi_0({\bf k})}|^2\delta(\omega - E_n + E_0)  \\
    +|\bra{\psi_0({\bf k})}c_{{\bf k \alpha \sigma }}\ket{\psi_m({\bf k})}|^2\delta(\omega +E_n-E_0)),
\end{multline}
for the spectral function.

As noted previously, computing $A(\omega)$ simply requires that we find the locus of intersection of the eigenregions, restricting our analysis to the low doping limit. The new quantity to be computed here is the integrand with the ground states given by $\ket{\psi_0} = \ket{E_1}, \ket{E_2}$. Since the relevant interested quantity is the spectral weight transfer, we want to integrate the spectral weight from $\omega=0$ to a low energy cutoff. To do so, we truncate the summation to the eigenbands that exceed the bandwidth in the region $0.5\leq s \leq 2$. We can also ignore the second term in the summand, since it is non-zero only for $\omega<0$. Thus, for the two cases of interest, $\ket{\psi_0} = \ket{E_1}, \ket{E_2}$, the integrand is given by
\begin{gather}\label{eq:swt_integrand}
    \sum_{n}(\bra{E_n}c_k\ket{E_1}\bra{E_1}c_k^\dagger \ket{E_n} = 0 \\
    \sum_{n}(\bra{E_n}c_k\ket{E_1}\bra{E_1}c_k^\dagger \ket{E_n} = 1+\frac{2t\eta({{\mathbf k}})}{\sqrt{U^2+64t^2\eta^2({\mathbf k})}}.
\end{gather}
Performing the integral over the Brillouin zone as before yields the full spectral weight transfer,
\begin{equation}
\begin{aligned}
    \Lambda(s) =& \left(1+\frac{4t}{\sqrt{U^2+256t^2}}\right)\frac{R^2}{4\pi}\\&+\frac{R^4tU^2}{4\pi(256t^2+U^2)^{3/2}}+\mathcal{O}(R^6).
\end{aligned}
\end{equation}
Rewriting the above in terms of the doping away from half filling, which is given by $x=n-1$, we obtain the series expansion,
\begin{gather}
    \Lambda(x) = 2x\left(1+\frac{4(t/U)}{\sqrt{1+256(t/U)^2}}\right) \\ +16\pi x^2\frac{(t/U)} {\left(1+256(t/U)^2\right)^{3/2}} + \mathcal{O}(x^3),
\end{gather}
for the spectral weight transfer as a function of the doping.
In particular, we can write the excess contribution to the spectral weight,
\begin{gather}
    f(t,x,U) = \frac{8t}{U} \frac{1}{\sqrt{1+256(t/U)^2}}x  \\ + 16\pi \frac{t}{U}\frac{1}{(1+256(t/U)^2)^{3/2}}x^2,
\end{gather}
laying plain the hopping dependence.
Thus the slope of the dynamic component of the spectral weight transfer is given simply as $4t/U$.  This will be confirmed by subsequent numerical calculations.

The above computation can be generalized to the case of arbitrary parameter values to obtain an analytic upper bound on the dynamic spectral weight transfer. Let us assume that the Brillouin zone is partitioned into sectors $\Omega_1$ and $\Omega_2$. The contribution to the LESW only comes from the $E_2(\bf k)$ band, corresponding to the $\Omega_2$ region. The SWT is then given by
\begin{gather}
    \Lambda(s) = \int_{\Omega_2}d{\bf k}\left(1+\frac{2t\eta({{\mathbf k}})}{\sqrt{U^2+64t^2\eta^2({\mathbf k})}}\right) \\
    \Lambda(s)\leq \left(1+\frac{4t}{\sqrt{U^2+64t^2}}\right)|\Omega_2|.
\end{gather}
Using Eq. \ref{eqn:occ_general}, we can rewrite the expression,
\begin{equation}
    \Lambda(x)\leq 2x\left(1+\frac{4t}{\sqrt{U^2+64t^2}}\right),
\end{equation}
 in terms of the doping.
This is in fact a fairly good estimate for the spectral weight at low doping, since the maximum is achieved at $\bf k = 0$ and the slope of the integrand is heavily damped by the large numerical factor in the denominator. 

\section{Symmetry about $1/4$ and $3/4$ filling}

\begin{figure}
    \centering
    \includegraphics[scale=0.31]{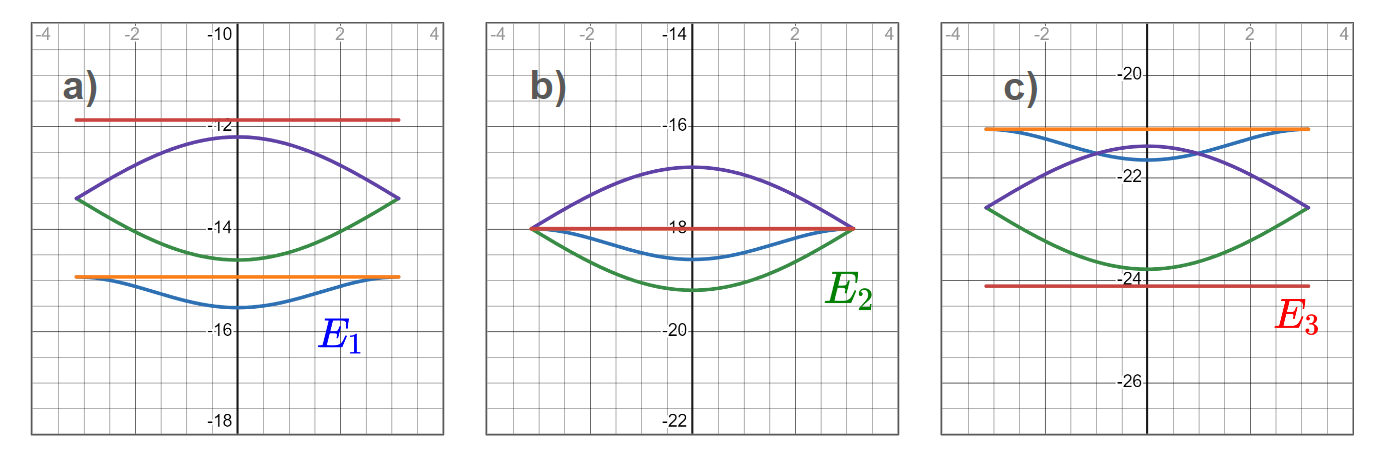}
    \caption{Plots of the eigenbands at different doping levels. The x-axis is the a slice along the Brillouin Zone with $k_y=0$. a) shows the bands at half filling when the only occupied band is the $N=2$ band labelled by $E_1$. b) Positions of bands for the symmetry point where $mu=U$ and the only occupied band has $N=3$ given by the $E_2$. c) Positions of bands in the fully filled system where $E_3$ is the only occupied band with $N=4$}
    \label{fig:DopedBands}
\end{figure}

The DSWT is symmetric as a function of the doping about the point $x=0.5$, which corresponds to $3/4$ filling. This is made apparent by the following argument. Consider $s\gg1$ and we are deep in the $N=2$ occupation. As $s$ is decreased, the locus of the intersection of $E_2$ and $E_3$, 
\begin{equation}
    E_2(k) = E_3(k),
\end{equation}
determines the occupation.
As before, we limit to low hole doping, allowing us to expand the cosine to quadratic order, which yields a circle of a different radius $R_*$. Thus, the occupation under hole doping is now given by
\begin{equation}
\label{n(s)}
    n(s) = 2-\frac{R_*^2}{2\pi}.
\end{equation}
Similarly, the spectral weight transfer reduces rto
\begin{equation}
    \Lambda(s) = 2 - (1+\frac{4t}{U})\frac{R_*^2}{\pi},
\end{equation}
or equivalently
\begin{equation}
    \Lambda(x) = 2 -2x(1+\frac{4t}{U}),
\end{equation}
which we obtain by eliminating $R^\ast$ through Eq. (\ref{n(s)}). Note that $x$ here is the hole doping fraction from the fully occupied state. 

The fillings of $1/4$ and $3/4$ occur at $\mu=0$ and $\mu=U$ respectively.  While we have shown that this symmetry obtains for the 2-orbital case, we anticipate that it obtains for even orbital number systems. To see this, note that the $1/4$ and $3/4$ filling points are where the dynamic contributions to the spectral weight are maximal. 

\section{Numerical Results}
\begin{figure}
    \centering
    \includegraphics[scale=0.35]{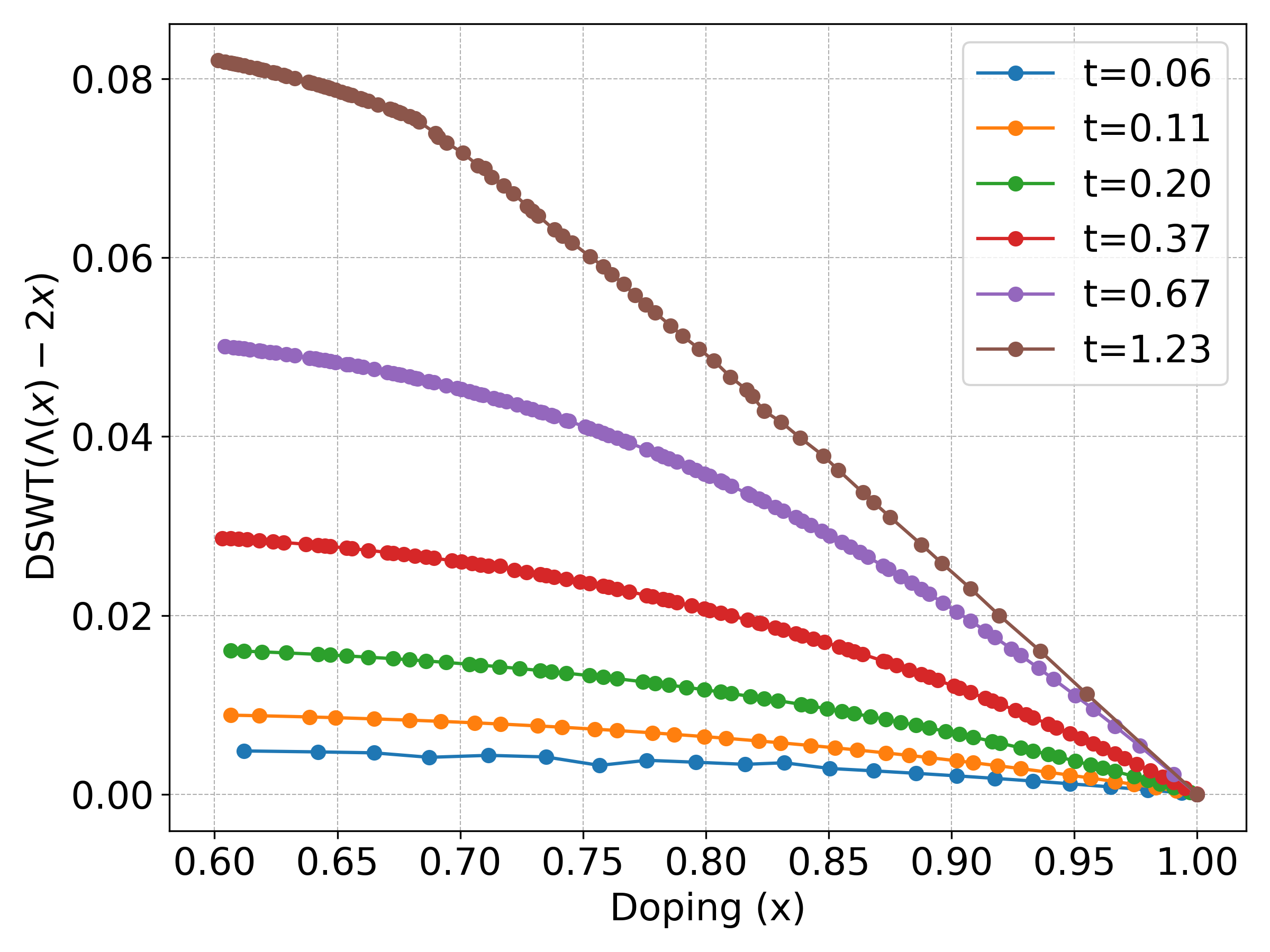}
    \caption{Dynamical spectral weight transfer $\Lambda(x)-2x$ for 2 orbital HK as a function of the doping.  Clearly shown is the excess of the DSWT beyond the static value of $2x$ which increases as the hopping increases.}
    \label{fig:dswt_2orb}
\end{figure}

The complexity of $N_\alpha$ orbital HK is identical to that of the $N_\alpha$-site Hubbard model. Thus, an analytic solution for large $N_\alpha$ is untenable at present. However, we can numerically diagonalize the 4-Orbital HK Hamiltonian to compute the spectral weight transfer (SWT). Subtracting off the static SWT gives us the dynamic part, shown in Fig. \ref{fig:dswt_2orb}.   As $t/U$ increases, the dynamical contribution to the DSWT increases, thereby laying plain the dynamical nature of this correction to the total spectral weight.  The increase of the spectral weight as a function of $t/U$ regardless of the doping appears to exhibit a universal rise.  To extract the slope at low doping, we perform a spline interpolation on the DSWT and analytically extract the slope at $x=1$. 

\begin{figure}
    \centering
    \includegraphics[scale=0.35]{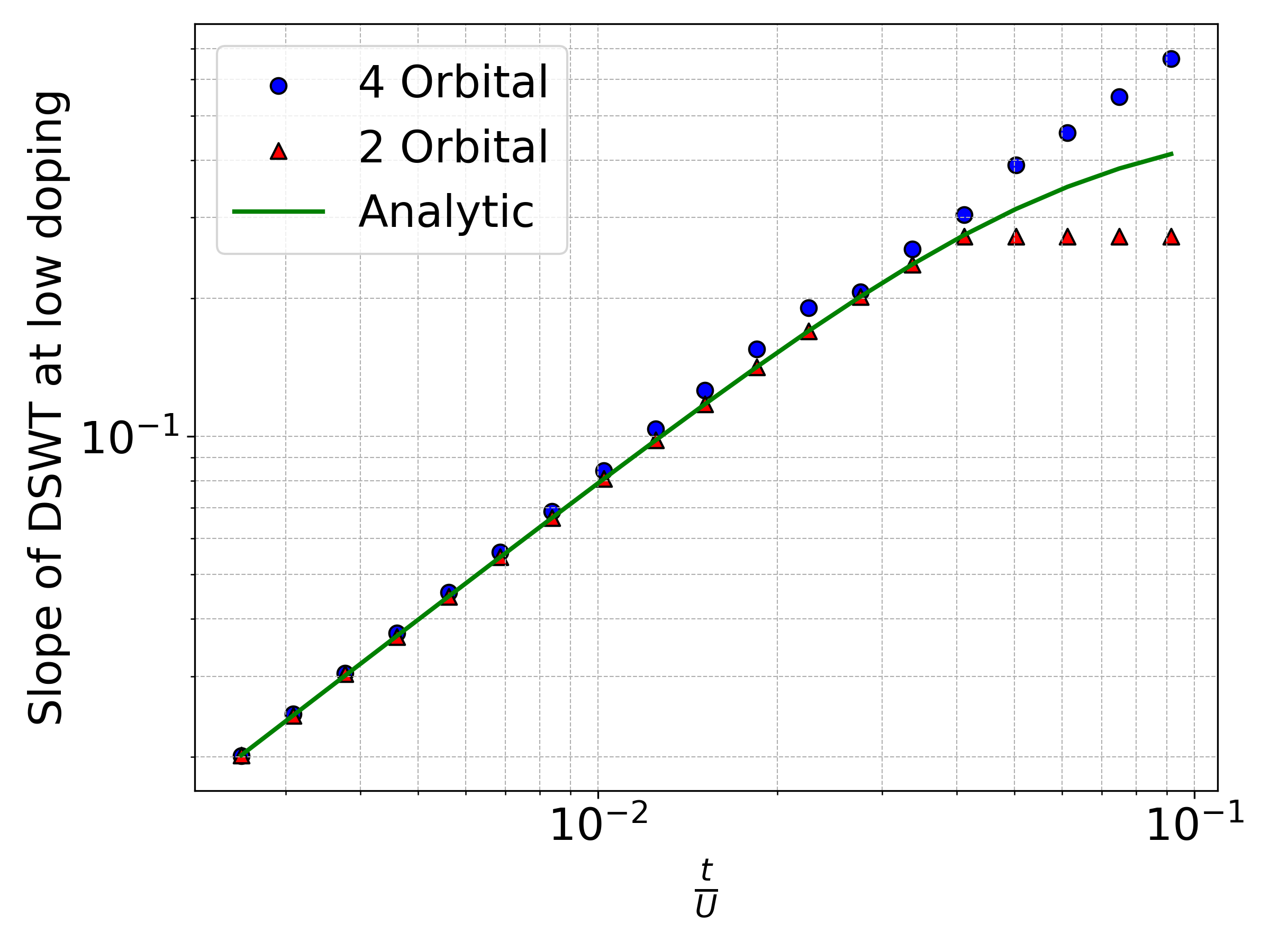}
    \caption{The slope of the DSWT extracted numerically for 2 and 4 Orbital HK in 2D. Observe that the results from the derived analytic expression diverge at the threshold $t/U = E^\ast$}
    \label{fig:slope_numeric}
\end{figure}

 The result is shown in Fig. \ref{fig:slope_numeric}.  What is evident here is that the slope is independent of the number of orbitals up to a critical value of $t/U$ that demarcates the transition to the weakly interacting regime. The critical value in general depends on the single-particle dispersion used for the model. At larger values of $t/U$, the band curvature results in a Brillouin Zone partition that is annular as shown in Fig. \ref{fig:BZpartition}. Thus, we cannot expand around the bottom of the band, which is necessary to evaluate the spectral weight integral. The independence of the slope on the number of orbitals underscores the fact that the OHK model is governed by an underlying fixed point\cite{fixedpoint} and hence no observable ultimately depends on the approach to the Hubbard model.

\begin{figure}
    \centering
    \includegraphics[scale=0.35]{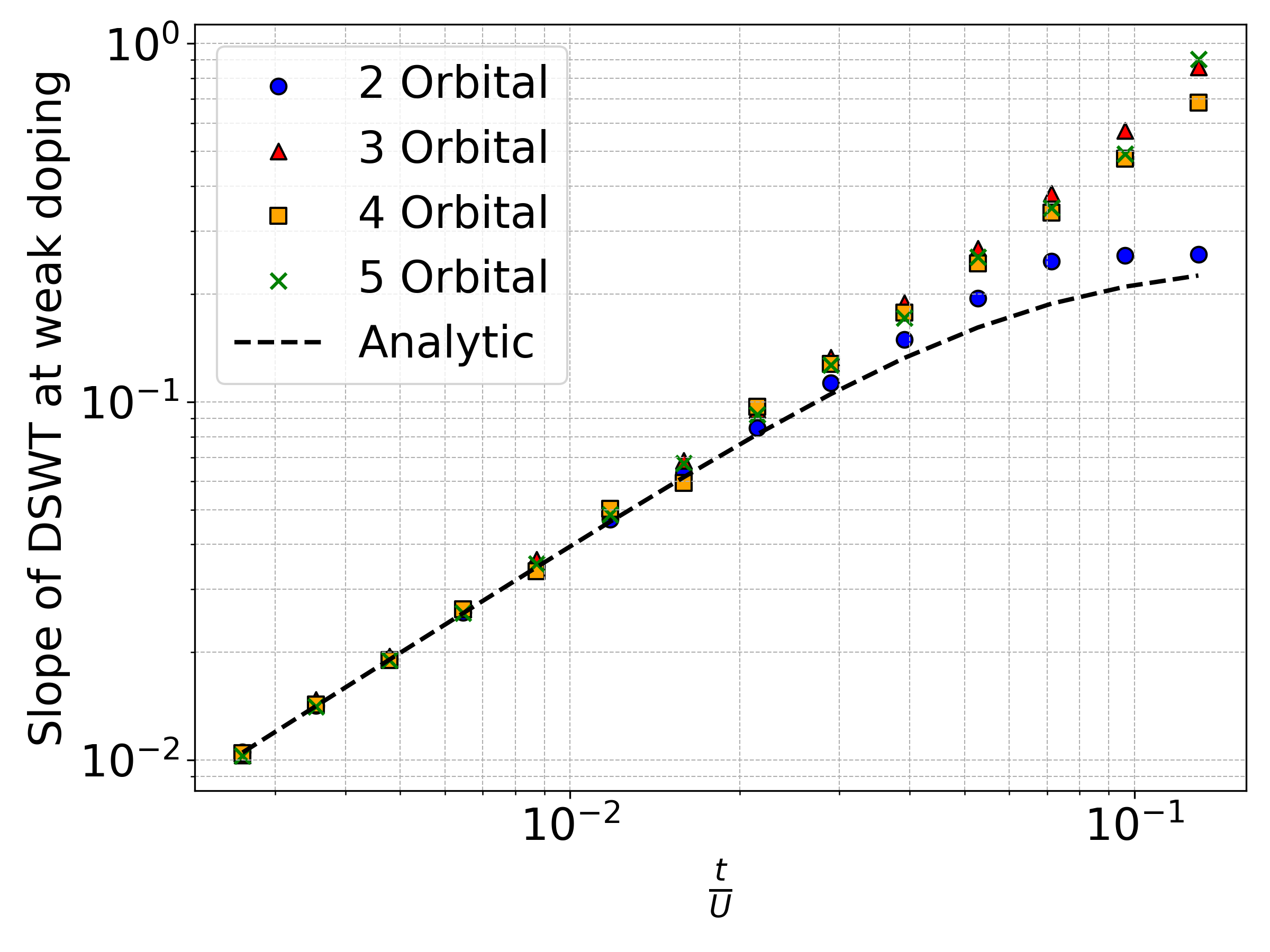}
    \caption{Numerically extracted slopes for Orbital HK in 1D. We observe a similar agreement at low $t/U$ across orbitals and an agreement with the analytic expression.}
    \label{fig:1d_slope_numeric}
\end{figure}

We observe a good fit to the analytic result for both the 2-orbital and the 4-orbital HK models below the identified threshold $t/U < 16\sqrt{3}$. 
 Recall, the threshold $16\sqrt{3}$ is the value at   which the intersection between the $N=3$ and $N=2$ eigenbands morphs from a circle to an annulus, as depicted in Fig. \ref{fig:BZpartition}.  Beyond this threshold, a well defined separation between the upper and lower bands ceases and no clear criterion permits a unique assignment of the LESW. Consequently, as long as the Hubbard bands are well separated, the slope of the LESW is independent or orbital number.  While numerics beyond 4-orbital in 2D are infeasible for large systems, we can repeat the same analysis for the 1D Orbital HK model, where we also have access to analytic results. In the Appendix, we derive a general result for 2-orbital HK in d-dimensions at low doping and at lowest order in $t/U$. For the 1D model, the single particle dispersion is $\eta(k) = \cos(k/2)$ and
\begin{gather}
    \Lambda(x) = 2x\left(1+2\frac{t}{U}\right).
\end{gather}
Observe that in 1D, this is consistent with the low doping results due to Eskes et al. \cite{eskes_analytic_1d},
\begin{equation}
    \Lambda(x) = 2x + \frac{4t}{\pi U}\text{sin}(\pi x) +\mathcal{O}((t/U)^2),
\end{equation}
which yields a slope of $4t/Ux$ in the low-doping limit.
Further, we find that at small $t/U$, the numerically obtained slope up to 5-orbital HK is consistent with the analytic result, as shown in Fig. \ref{fig:1d_slope_numeric}.  This amplifies that the orbital index $n$ plays no role in the LESW.  Whether this remains the case for all experimental observable remains to be seen.

\section{Final Remarks}
While DSWT is well documented both experimentally\cite{chen} and theoretically\cite{sawatzky,eskes,eskes_analytic_1d}, an analytical treatment has been lacking.    We have shown that the OHK model is amenable to an analytical extraction of the DSWT. While the model approaches the Hubbard model in the infinite orbital limit, we find that the slope of the DSWT as a function of $t/U$ is independent of the orbital number.  The implication here is two-fold.  First, it indicates a rapid convergence to the Hubbard model.  Second, the DSWT is a generic feature of Mott physics determined entirely by the dimensionless ratio $t/U$ and not the orbital number. Since, OHK preserves the decoupling of the momenta as the model can still be written as $\sum_k h_k$, the dynamics are still governed by the HK fixed point\cite{fixedpoint}.  Hence, no experimental observable can depend on flow parameter that yields the Hubbard model, namely $N_{\alpha}$. Since the OHK model is exactly solvable for small numbers of orbitals, this platform affords an exact treatment of spectral weights in strongly correlated matter.  Consequently, the OHK model is an effective simulator of Mott physics described by the Hubbard model.

\section{Acknowledgements}
We acknowledge NSF DMR-2111379 for partial funding of this project.
This research was supported in part by the Illinois Computes project which is supported by the University of Illinois Urbana-Champaign and the University of Illinois System. We would also like to thank Peizhi Mai for useful discussions. 
\section{Appendix}
\subsection{DSWT in d-dimensions}
Our work here generalizes to any dimension.  The key to deriving the d-dimensional result is recognizing that the eigensystem at each $\bf{k}$ point is quite general for 2-orbital HK. The only differences arise from the dimensionality of the Brillouin Zone and the single particle dispersion $\eta({\mathbf{k}})$. 

To proceed, we note that the general result for the occupation as a function of the chemical potential is the same as before,
\begin{equation}
    n(s) = |\Omega^d_1| + \frac32 |\Omega^d_2|,
\end{equation}
where now $\Omega^d_1$ and $\Omega^d_2$ are in general higher dimensional regions. In the low doping and flat-band limit, $\Omega^d_2$ is a hypersphere of some radius $R$. We use the fact that the Brillouin Zone has total volume $(2\pi)^d$ to obtain,
\begin{equation}
    n(s) = 1+\frac{V_d(R)}{2(2\pi)^d},
\end{equation}
where $V_d(R)$ is the volume of a d-dimensional ball with radius $R(s)$. Next, we use the results in Eq. \ref{eq:swt_integrand} to compute the spectral weight transfer as,
\begin{equation}
    \Lambda(s) = \frac{1}{(2\pi)^d}\int_{\mathbb{B}_R}d^d \mathbf{k} \left(1+\frac{2t\eta({\mathbf{k}})}{\sqrt{U^2+64t^2\eta^2({\mathbf{k}})}}\right).
\end{equation}
To first order in the radius $R$, we obtain the expression,
\beq
    \Lambda(s) = \frac{V_d(R)}{(2\pi)^d}(1+2t\eta(\mathbf{k_0})) \\
    \implies \Lambda(x) = 2x\left(1+2\eta(\mathbf{k_0})\frac{t}{U}\right),
\eeq
as the general $d-$dimensional expression for the spectral weight. Note that the $d-$dependence is entirely contained in the single particle dispersion $\eta(\mathbf{k})$ at $\mathbf{k_0} = \mathbf{0}$. While this is in general dependent on how the unit cells are defined, if we systematically construct the $d-$dimensional system by stacking $(d-1)-$dimensional systems and shifting each adjacent system by a lattice spacing, we arrive at a general expression:
\begin{equation}
    \eta)\mathbf{k}) = \sum_{i=1}^{d} \text{cos}\left(\frac{k_i}{2}\right).    
\end{equation}
This yields the universal result
\begin{equation}
    \Lambda(x) = 2x+4d\frac{t}{U}x.
\end{equation}
\bibliography{dswt}
\end{document}